# NMR Dynamics of Quantum Discord for Spin-Carrying Gas Molecules in a Closed Nanopore

**M. A. Yurishchev**

*Institute of Problems of Chemical Physics, Russian Academy of Sciences, Chernogolovka, Moscow oblast, 142432 Russia*
*e-mail: yur@itp.ac.ru*

**Abstract**—A local orthogonal transformation that transforms any centrosymmetric density matrix of a two-qubit system to the X form has been found. A piecewise-analytic–numerical formula $Q = \min\{Q_{\pi/2}, Q_\theta, Q_0\}$, where $Q_{\pi/2}$ and $Q_0$ are analytical expressions and the branch $Q_\theta$ can be obtained only by numerically searching for the optimal measurement angle $\theta \in (0, \pi/2)$, is proposed to calculate the quantum discord $Q$ of a general X state. The developed approaches have been applied for a quantitative description of the recently predicted flickering (periodic disappearance and reappearance) of the quantum-information pair correlation between nuclear 1/2 spins of atoms or molecules of a gas (for example, $^{129}$Xe) in a bounded volume in the presence of a strong magnetic field.

## 1. INTRODUCTION

The notion of quantum discord introduced to describe all purely quantum correlations in a system currently occupies a central place in discussing the questions of radically increasing the performance of computers, creating secure data transmission networks, implementing the teleportation of states, detecting quantum phase transitions in condensed matter, etc. The progress achieved in the theory and applications of quantum discord is reflected in the recent reviews [1–3].

The actual calculation of quantum discord and, hence, the investigation of the behavior of quantum correlations in various physical systems is severely limited by the necessity of performing the optimization procedure entering into the definition of this quantity: the problem belongs to the complete NP class of complexity [4]. Even for two-qubit models, where the density matrix has sizes of only 4 × 4, the achievements by analytical methods are exhausted by the special cases of X density matrices, the Bell-diagonal states (include three independent parameters) [5]. (The term "X state" was introduced in 2007 and means a fourth-order density matrix in which only the diagonal and antidiagonal entries can be nonzero [6, 7]; an arbitrary X density matrix contains seven real parameters.) Attempts were also made to find analytical formulas for calculating the quantum discord of more general families of X density matrices [8–11]. However, in [12–14] the cases where the analytical formulas [8–11] are incorrect were demonstrated with specific examples, and it was concluded that "for general two-qubit X states quantum discord cannot be evaluated analytically" [14]. In this paper, we will show that the general formula for calculating the quantum discord of X density matrices is actually piecewise-analytic–numerical, i.e., the discord is expressed analytically in some regions of the domain of definition and only numerically in other regions.

Let us now turn to the physical problem of interest to us. The behavior of gaseous hydrogen in elongated microscopic voids of amorphous silicon was investigated by nuclear magnetic resonance (NMR) methods in [15]. For this purpose, hydrogen was pumped into silicon films with a thickness of two microns under a pressure of 2 kbar. Subsequently, the produced a-Si:H sample was studied in an NMR spectrometer with a magnetic field of 4.7 T. The experimental results turned out to be unexpected in many respects. The gas behaved as a quantum collective system in which all nuclear spins of orthohydrogen interacted with one another in the same way, irrespective of their positions in cavity space and, consequently, the distances between them. The authors of [15] (see also [16]) interpreted their observations, first, by rapid thermal molecular motion in the nanocavity in combination with relatively slow spin-flip rates and, second, by an asymmetry (nonsphericity) of the voids. As a result, these factors led to an incomplete averaging of the secular part of the magnetic dipole–dipole interactions between particle spins. Subsequently, the problem of the quantum behavior of spin-carrying particles in a closed volume of space in the presence of a strong magnetic field was considered in [17–21].

Previously [22], we made an attempt to theoretically study the dynamical quantum correlations in a model of $N$ moving $I = 1/2$ spins confined in the presence of a strong magnetic field $B_0$ in a volume with an ellipsoidal shape. A two-particle density matrix in a rotating reference frame was found within the framework of the theory of an averaged Hamiltonian with a universal (the same for all pairs of particles) coupling constant $D$. (The quantum correlations in a rotating reference frame are also of interest in the problems of the behavior of atoms in the field of an optical cavity [23]). As it turned out, the density matrix for pairs of particles in a nanopore has not the X form but the cetrosymmetric (CS) one. This allowed us to study only the quantum entanglement in the system but did not allow the quantum discord to be investigated in full. (The CS $n \times n$ matrix is defined by the relations $a_{i,j} = a_{n-i+1, n-j+1}$ for its matrix elements; the properties of CS matrices are outlined in [24–26].) Fortunately, the density matrix found acquires a three-parameter X form at the instants of time $t_l = \pi l/3D$ ($l = 0, 1, 2, ...$) and in the intervals between these instants, but in the thermodynamic limit $N \rightarrow \infty$. We calculated the quantum discord for these two special cases. It emerged that the quantum discord becomes zero at the instants of time $t_l$ and has a finite value in the intervals between them. Such periodic disappearances of correlations and their reappearances were called discord flickering [22]. At the same time, the complete picture of the behavior of quantum discord remained unknown.

In this paper, we find, first, a universal local orthogonal similarity transformation that transforms any fourth-order CS matrix into an X matrix (and vice versa). Since the quantum discord is invariant relative to local unitary transformations [1–3], this finding allows the problem of calculating the discord of a CS state to be reduced the analogous problem for an X state. Second, we calculate the quantum discord of the X density matrix using a piecewise-analytic–numerical formula, which guarantees that the calculation of the discord function is correct in its entire domain of definition.

In Sections 2–7, we give a general definition of quantum discord, describe the model and present the density matrix for pairs of spin-carrying particles in a nanopore, find the relationship between the CS and X matrices via a local orthogonal transformation, derive the piecewise-analytic–numerical formula for calculating the discord of an arbitrary X state, and present and discuss the results for the quantum-information correlation between pairs of particles in a nanopore. In the Conclusions, we briefly summarize the results of our study.

## 2. CORRELATIONS AND THEIR MEASURES

In the theory of probability and mathematical statistics, the correlation between two random variables $x$ and $y$ with a joint probability distribution function $p(x, y)$ is measured by the covariance (centered correlation moment)

$$\mathrm{cov}(x, y) = \overline{(x - \bar{x})(y - \bar{y})}, \quad (1)$$

where the overbar denotes an averaging over the probability distribution. Unfortunately, zero covariance does not yet guarantee the independence of the random variables, i.e., that $p(x, y) = p_1(x)p_2(y)$ [27–29].

On the other hand, the so-called mutual information plays a major role in the classical information theory [30–33]:

$$\mathcal{I}(X:Y) = H(X) + H(Y) - H(X, Y), \quad (2)$$

where $H(X)$, $H(Y)$, and $H(X, Y)$ are the Shannon entropies. It turned out that, in contrast to the covariance, zero mutual information, $\mathcal{I} = 0$, is a necessary and sufficient condition for the independence of $X$ and $Y$. This allows $\mathcal{I}$ to be used as a measure of the now information correlation (statistical relations) between systems $X$ and $Y$ [29]. This measure corresponds to an intuitive expectation: two objects or phenomena correlate (are interdependent) if in one of them there is a fraction of information about the other.

Using the Bayes formula known in the probability theory, the right part of equality (2) for the mutual information can be written in Shannon's nonsymmetric form that we will denote by $\mathcal{J}$:

$$\mathcal{J}(X:Y) = H(X) - H(X|Y), \quad (3)$$

where

$$H(X|Y) = H(X, Y) - H(Y) \quad (4)$$

is the conditional entropy. In the classical case, of course, $\mathcal{J} \equiv \mathcal{I}$.

In the quantum information theory [34–36], Eq. (2) is replaced by the new relation

$$\mathcal{I}(A:B) = S(\rho_A) + S(\rho_B) - S(\rho_{AB}), \quad (5)$$

which serves as a measure of the mutual information between two subsystems $A$ and $B$ comprising together the joint system $AB = A \cup B$. In Eq. (5), $\rho_{AB}$ is the density matrix of the joint system $AB$, $\rho_A$ and $\rho_B$ are the reduced density matrices for subsystems $A$ and $B$, respectively, and $S(\rho)$ ($\rho = \{\rho_A, \rho_B, \rho_{AB}\}$ is the von Neumann entropy,

$$S(\rho) = -\mathrm{Tr}\rho\log\rho. \quad (6)$$

(The choice of the logarithm base fixes the measurement unit for the information entropy: nats, bits, dits, hartleys, bans, etc.). It is important that, as in the classical case for the distribution function, the quality $\mathcal{I} = 0$ is a necessary and sufficient condition for the factorization of the density matrix of the joint system: $\rho_{AB} = \rho_A \otimes \rho_B$, implying that $A$ and $B$ are absolutely independent (uncorrelated) (see, e.g., [36]). In the quantum information theory, $\mathcal{I}$ serves as a measure of all (both classical $C$ and quantum $Q$) correlations between two (sub)systems: $\mathcal{I} = C + Q$.

Ollivier and Zurek [37] generalized Shannon's relation (3) to the quantum case:

$$\mathcal{J}(A:B) = S(\rho_A) - \sum_i p_i S(\rho_A^i), \quad (7)$$

where the second term in the right-hand side of this equality represents the quantum conditional entropy. In addition,

$$\rho_A^i = \text{Tr}_B(B_i \rho_{AB} B_i^+)/\text{Tr}_{AB}(B_i \rho_{AB} B_i^+) \quad (8)$$

is the state of subsystem $A$ that arises once the measurement has been made on subsystem $B$ and it has been recorded that it gave result $i$ (a classical outcome), and

$$p_i = \text{Tr}_{AB}(B_i \rho_{AB} B_i^+) \quad (9)$$

is the probability to obtain result $i$ after the measurement. The minimal difference $\mathcal{I} - \mathcal{J} \equiv Q$ that is reached on the set of all one-dimensional projective measurements $\{B_i\}$ (von Neumann measurements) was identified with a measure of the quantumness of correlations and was called quantum discord—discord between $\mathcal{I}$ and $\mathcal{J}$, between classicality and quantumness [37]. Thus, the quantum discord is defined by the equality

$$Q = S(\rho_B) - S(\rho_{AB}) + \min_{\{B_i\}} \sum_i p_i S(\rho_A^i). \quad (10)$$

## 3. HAMILTONIAN AND TWO-PARTICLE DENSITY MATRIX

In a rotating reference frame, the effective Hamiltonian for $N$ nuclear 1/2 spins of gas (or liquid) molecules subjected to strong thermal motions in an axially symmetric ellipsoidal container is ([22] and references therein)

$$H_{dz} = \frac{D}{2}(3I_z^2 - \mathbf{I}^2), \quad (11)$$

where $D$ is the coupling constant dependent on the shape of the cavity (the ellipsoid axis ratio), its volume $V$, and its orientation (inclination) with respect to the external magnetic field; $I_z$ is the sum of the projections of the spin angular momenta for all particles onto the $z$ axis (the direction of the external magnetic field); $\mathbf{I}^2$ is the square of the total spin angular momentum for the particles. Note that $D \propto 1/V$ and, therefore, the spin coupling actually manifests itself only for nanosized cavities. In addition, the couplings are averaged to zero in the case of a spherically symmetric cavity: $D = 0$ [15, 17].

As in [22], here we will assume that the sample in the NMR spectrometer is initially brought to a state of complete thermal equilibrium. Then, at the instant of time $t = 0$, a single short resonance $\pi/2$-pulse turning all spins around the $y$ axis through an angle of $\pi/2$ is applied to it. The system's density matrix takes a non-equilibrium form:

$$\rho_N(0) = \frac{1}{Z}\exp(\beta I_x). \quad (12)$$

Thereafter, the process of free evolution begins. It is described, in accordance with the Liouville equation, by the density matrix

$$\rho_N(t) = \exp(-iH_{dz}t)\rho_N(0)\exp(iH_{dz}t)$$
$$= \frac{1}{Z}\exp(-i\alpha I_z^2)\exp(\beta I_x)\exp(i\alpha t I_z^2). \quad (13)$$

In these equalities,

$$Z = \text{Tr}\{\exp(\beta I_x)\} = 2^N \cosh^N \frac{\beta}{2} \quad (14)$$

is the system's partition function, $\beta = \hbar\omega_0/k_B T$ is the inverse dimensionless temperature, $\omega_0 = \gamma B_0$ ($\gamma$ is the gyromagnetic ratio) is the Larmor frequency, $T$ is the temperature, and $\alpha = 3D/2$.

Previously [22], we showed that the density matrix for any two chosen spins ($A$ and $B$) obtained by reducing $\rho_N$ over the remaining $N - 2$ spins is

$$\rho_{AB} = \begin{pmatrix} \frac{1}{4} & \frac{1}{2}p - iu & \frac{1}{2}p - iu & q - r \\ \frac{1}{2}p + iu & \frac{1}{4} & q + r & \frac{1}{2}p + iu \\ \frac{1}{2}p + iu & q + r & \frac{1}{4} & \frac{1}{2}p + iu \\ q - r & \frac{1}{2}p - iu & \frac{1}{2}p - iu & \frac{1}{4} \end{pmatrix}, \quad (15)$$

where

$$p = \frac{1}{2}\tanh\frac{\beta}{2}\cos^{N-1}(\alpha t),$$
$$q = \frac{1}{8}\tanh^2\frac{\beta}{2}[1 + \cos^{N-2}(2\alpha t)],$$
$$r = \frac{1}{8}\tanh^2\frac{\beta}{2}[1 - \cos^{N-2}(2\alpha t)], \quad (16)$$
$$u = \frac{1}{4}\tanh\frac{\beta}{2}\cos^{N-2}(\alpha t)\sin(\alpha t).$$

Expanding the density operator (15) in the Pauli matrices $\sigma_i^\nu$ ($\nu = x, y, z$ and $i = 1, 2$) brings $\rho_{AB}$ to the Bloch form

$$\rho_{AB} = \frac{1}{4}[1 + 2p(\sigma_1^x + \sigma_2^x) + 4q\sigma_1^x\sigma_2^x$$
$$+ 4r\sigma_1^y\sigma_2^y + 4u(\sigma_1^y\sigma_2^z + \sigma_1^z\sigma_2^y)]. \quad (17)$$

The state (15) has no X form and, as we know, there are no tools for calculating its quantum discord in the literature. In this paper, however, we give a method for such a calculation. Our approach is based primarily on revealing the relationship between the CS and X matrices. We now turn to considering this question.

## 4. RELATIONSHIP BETWEEN THE CS AND X MATRICES

The fourth-order CS matrix can be written in the most general form as

$$A = \begin{pmatrix} a_1 & a_2 & a_3 & a_4 \\ a_5 & a_6 & a_7 & a_8 \\ a_8 & a_7 & a_6 & a_5 \\ a_4 & a_3 & a_2 & a_1 \end{pmatrix}, \quad (18)$$

where $a_1, ..., a_8$ are any real or complex quantities. Let us subject this matrix to the similarity transformation $H_2 A H_2$, in which

$$H_2 = H \otimes H = \frac{1}{2}\begin{pmatrix} 1 & 1 & 1 & 1 \\ 1 & -1 & 1 & -1 \\ 1 & 1 & -1 & -1 \\ 1 & -1 & -1 & 1 \end{pmatrix}, \quad (19)$$

where

$$H = \frac{1}{\sqrt{2}}\begin{pmatrix} 1 & 1 \\ 1 & -1 \end{pmatrix} \quad (20)$$

is the ordinary Hadamard transform. As a result, we obtain the matrix

$$H \otimes H A H \otimes H = \begin{pmatrix} b_1 & & & b_2 \\ & b_3 & b_4 & \\ & b_5 & b_6 & \\ b_7 & & & b_8 \end{pmatrix}, \quad (21)$$

with an X structure (resembling the letter "X"). The nonzero matrix elements $b_1, ..., b_8$ in Eq. (21) are

$$\begin{aligned}
b_1 &= (a_1 + a_2 + a_3 + a_4 + a_5 + a_6 + a_7 + a_8)/2, \\
b_2 &= (a_1 - a_2 - a_3 + a_4 + a_5 - a_6 - a_7 + a_8)/2, \\
b_3 &= (a_1 - a_2 + a_3 - a_4 - a_5 + a_6 - a_7 + a_8)/2, \\
b_4 &= (a_1 + a_2 - a_3 - a_4 + a_5 + a_6 - a_7 - a_8)/2, \\
b_5 &= (a_1 - a_2 + a_3 - a_4 + a_5 - a_6 + a_7 - a_8)/2, \\
b_6 &= (a_1 + a_2 - a_3 - a_4 + a_5 + a_6 - a_7 - a_8)/2, \\
b_7 &= (a_1 + a_2 + a_3 + a_4 - a_5 - a_6 - a_7 - a_8)/2, \\
b_8 &= (a_1 - a_2 - a_3 + a_4 - a_5 + a_6 + a_7 - a_8)/2.
\end{aligned} \quad (22)$$

It is easy to verify that, conversely, the double Hadamard transform (19) brings any X matrix to the CS form.

The key point is that the transformation (19) is local and orthogonal. As applied to Hermitian matrices, the ability of $H_2$ to transform CS $\longleftrightarrow$ X was first established in our preprint [38]. Here, this property was extended to CS and X matrices of a general form.

The revealed relationship between the CS and X matrices opens the fundamental possibility of calculating the quantum discord for particles in a nanopore.

## 5. THE REAL X FORM OF A TWO-PARTICLE DENSITY MATRIX

In this section, we will make a series of local unitary transformations that will bring the CS matrix (15) to a form suitable for performing the discord calculations. Since under such transformations we will be dealing basically with the same matrix but only in different representations, we will retain the symbol $\rho_{AB}$ for it in order not to introduce new designations.

First of all, we will transform $\rho_{AB}$ to the X form by applying the method developed in the previous section:

$$H \otimes H \rho_{AB} H \otimes H \longrightarrow \rho_{AB}$$

$$= \begin{pmatrix} \frac{1}{4} + p + q & 0 & 0 & -r + 2iu \\ 0 & \frac{1}{4} - q & r & 0 \\ 0 & r & \frac{1}{4} - q & 0 \\ -r - 2iu & 0 & 0 & \frac{1}{4} - p + q \end{pmatrix}, \quad (23)$$

or in the Bloch form

$$\rho_{AB} = \frac{1}{4}[1 + 2p(\sigma_1^z + \sigma_2^z) + 4r\sigma_1^y \sigma_2^y \\ + 4q\sigma_1^z \sigma_2^z - 4u(\sigma_1^x \sigma_2^y + \sigma_1^y \sigma_2^x)]. \quad (24)$$

We see that the original Bloch expansion (17) transforms into (24) in accordance with the fact that the Hadamard transform (20) transforms the Pauli matrix $\sigma_x$ to $\sigma_z$, reverses the sign for the matrix $\sigma_y$, and transforms $\sigma_z$ to $\sigma_x$.

It is well known (see, e.g., [14]) that any X density matrix can be brought to a real form with nonnegative off-diagonal elements with retention of the X form using local rotations around the $z$ axis through appropriate angles. In our case, the complex parts of nondiagonal elements are removed by the rotation,

$$U = \exp\left(-\frac{i\varphi\sigma_z}{2}\right) \otimes \exp\left(-\frac{i\varphi\sigma_z}{2}\right), \quad (25)$$

of both spins through the same angle

$$\varphi = -\frac{1}{2}\arctan\frac{2u}{r}. \quad (26)$$

As a result,

$$U\rho_{AB}U^\dagger \longrightarrow \rho_{AB} = \begin{pmatrix} \frac{1}{4}+p+q & 0 & 0 & 2u\sin 2\varphi - r\cos 2\varphi \\ 0 & \frac{1}{4}-q & r & 0 \\ 0 & r & \frac{1}{4}-q & 0 \\ 2u\sin 2\varphi - r\cos 2\varphi & 0 & 0 & \frac{1}{4}-p+q \end{pmatrix}, \quad (27)$$

i.e., the Bloch form

$$\rho_{AB} = \frac{1}{4}[1 + 2p(\sigma_1^z + \sigma_2^z) + 4(r\sin^2\varphi + u\sin 2\varphi)\sigma_1^x\sigma_2^x \\ + 4(r\cos^2\varphi - u\sin 2\varphi)\sigma_1^y\sigma_2^y + 4q\sigma_1^z\sigma_2^z] \quad (28)$$

now acquired a normal, canonical form (without any cross terms like $xy$).

According to (16), $r \geq 0$. If another off-diagonal element of matrix (27) is negative, then, as is easy to verify, its sign can be easily reversed by the transformation

$$U_1 = \exp\left(-i\frac{\pi}{4}\sigma_z\right) \otimes \exp\left(i\frac{\pi}{4}\sigma_z\right) = \begin{pmatrix} i & & & \\ & 1 & & \\ & & 1 & \\ & & & -i \end{pmatrix}. \quad (29)$$

(If necessary, a similar transformation, $\exp(i\pi\sigma_z/4) \otimes \exp(-i\pi\sigma_z/4)$, can selectively act on the sign of the second off-diagonal element $r$.) In other words, once the X density matrix has been brought to a real form, its off-diagonal elements can be put under the absolute value sign and the magnitude of the quantum correlations in the system will not change.

As a result, for the two-particle density matrix we have

$$\rho_{AB} = \begin{pmatrix} \frac{1}{4}+p+q & 0 & 0 & |2u\sin 2\varphi - r\cos 2\varphi| \\ 0 & \frac{1}{4}-q & r & 0 \\ 0 & r & \frac{1}{4}-q & 0 \\ |2u\sin 2\varphi - r\cos 2\varphi| & 0 & 0 & \frac{1}{4}-p+q \end{pmatrix}. \quad (30)$$

## 6. THE PIECEWISE-ANALYTIC–NUMERICAL FORMULA

Thus, the general X density matrix

$$\rho_{AB} = \begin{pmatrix} a & 0 & 0 & u_1+iu_2 \\ 0 & b & v_1+iv_2 & 0 \\ 0 & v_1-iv_2 & c & 0 \\ u_1-iu_2 & 0 & 0 & d \end{pmatrix} \quad (31)$$

can always be brought to a real form with nonnegative matrix elements using local unitary transformations:

$$\rho_{AB} = \begin{pmatrix} a & 0 & 0 & |u| \\ 0 & b & |v| & 0 \\ 0 & |v| & c & 0 \\ |u| & 0 & 0 & d \end{pmatrix}. \quad (32)$$

In these equalities, $a + b + c + d = 1$,

$$a, b, c, d \geq 0, \quad ad \geq u_1^2 + u_2^2, \quad bc \geq v_1^2 + v_2^2 \quad (33)$$

(a consequence of the requirement that the definiteness of the density matrix be nonnegative), and

$$u = u_1\cos\left(\arctan\frac{u_2}{u_1}\right) + u_2\sin\left(\arctan\frac{u_2}{u_1}\right), \quad (34)$$

$$v = v_1 \cos\left(\arctan \frac{v_2}{v_1}\right) + v_2 \sin\left(\arctan \frac{v_2}{v_1}\right). \quad (35)$$

(In this section, we use the quantity $u$ that should not be confused with the pair correlator $u$ from Eqs. (16) for a nanopore.) Note that the original X density matrix (31) contained seven parameters, while in its real form (32) their number was reduced to five. Of course, this simplifies the solution of the problem of calculating the discord.

For a real X density matrix with nonnegative off-diagonal elements (actually, only the condition $uv \geq 0$ will suffice [39]), the measurements that enter into the definition of the quantum discord (10) are reduced to projections of the spin for subsystem $B$ onto the $z$ axis, with the optimal projection results being a function of only the polar angle $\theta$ (see [14] and references therein). As a consequence, the quantum discord is

$$Q = S(\rho_B) - S(\rho_{AB}) + \min_\theta S_{\text{cond}}(\theta), \quad (36)$$

where $\theta \in [0, \pi/2]$ is the angle that the projector makes with the $z$ axis when measuring the conditional entropy $S_{\text{cond}}$.

The entropy (in nats) of subsystem $B$ is

$$S(\rho_B) = -(a+c)\ln(a+c) - (b+d)\ln(b+d), \quad (37)$$

while the entropy of the joint system $AB$ is

$$S(\rho_{AB}) \equiv S$$

$$= -\frac{a+d+\sqrt{(a-d)^2+4u^2}}{2} \ln \frac{a+d+\sqrt{(a-d)^2+4u^2}}{2}$$

$$-\frac{a+d-\sqrt{(a-d)^2+4u^2}}{2} \ln \frac{a+d-\sqrt{(a-d)^2+4u^2}}{2} \quad (38)$$

$$-\frac{b+c+\sqrt{(b-c)^2+4v^2}}{2} \ln \frac{b+c+\sqrt{(b-c)^2+4v^2}}{2}$$

$$-\frac{b+c-\sqrt{(b-c)^2+4v^2}}{2} \ln \frac{b+c-\sqrt{(b-c)^2+4v^2}}{2}.$$

The quantum conditional entropy of subsystem $A$ once the measurement on $B$ has been performed is [14]

$$S_{\text{cond}}(\theta) = \Lambda_1 \ln \Lambda_1 + \Lambda_2 \ln \Lambda_2 - \sum_{i=1}^{4} \lambda_i \ln \lambda_i, \quad (39)$$

where

$$\Lambda_{1,2} = \frac{1}{2}[1 \pm (a-b+c-d)\cos\theta], \quad (40)$$

$$\lambda_{1,2} = \frac{1}{4}[1 + (a-b+c-d)\cos\theta$$

$$\pm \{[a+b-c-d+(a-b-c+d)\cos\theta]^2 \quad (41)$$

$$+ 4(|u|+|v|)^2 \sin^2\theta\}^{1/2}],$$

$$\lambda_{3,4} = \frac{1}{4}[1 - (a-b+c-d)\cos\theta$$

$$\pm \{[a+b-c-d-(a-b-c+d)\cos\theta]^2 \quad (42)$$

$$+ 4(|u|+|v|)^2 \sin^2\theta\}^{1/2}].$$

Obviously, the conditional entropy $S_{\text{cond}}(\theta)$ is a differentiable function of its argument $\theta$. In addition, as is easy to verify, its derivative $S'_{\text{cond}}(\theta)$ at the boundary points $\theta = 0$ and $\pi/2$ is zero: $S'_{\text{cond}}(0) = S'_{\text{cond}}(\pi/2) = 0$.

Equations (37)–(42) allow the non-optimized (measurement-dependent) discord to be defined as

$$Q(\theta) = S(\rho_B) - S(\rho_{AB}) + S_{\text{cond}}(\theta). \quad (43)$$

The absolute minimum of this function can be either at the boundary points ($\theta = 0$ or $\pi/2$) or between these points: $\theta \in (0, \pi/2)$. Thus, there are a total of three alternatives for calculating the quantum discord:

$$Q = \min\left\{Q_0, Q_\theta, Q_{\frac{\pi}{2}}\right\}. \quad (44)$$

Here,

$$Q_0 \equiv Q(0) = -S - a\ln a - b\ln b - c\ln c - d\ln d, \quad (45)$$

$$Q_{\frac{\pi}{2}} \equiv Q(\pi/2) = -S - \ln 2 - (a+c)$$

$$\times \ln(a+c) - (b+d)\ln(b+d)$$

$$-\frac{1+\sqrt{(a+b-c-d)^2+4(|u|+|v|)^2}}{2} \quad (46)$$

$$\times \ln \frac{1+\sqrt{(a+b-c-d)^2+4(|u|+|v|)^2}}{4}$$

$$-\frac{1-\sqrt{(a+b-c-d)^2+4(|u|+|v|)^2}}{2}$$

$$\times \ln \frac{1-\sqrt{(a+b-c-d)^2+4(|u|+|v|)^2}}{4}.$$

Equality (44) generalizes the previously used relation [5, 8–11]

$$Q = \min\{Q_0, Q_{\pi/2}\}, \quad (47)$$

the latter is valid if either only $\sigma_z$ or only $\sigma_x$ is an optimal observable, while the branch

$$Q_\theta = \min_{\theta \in (0, \pi/2)} Q(\theta) \quad (48)$$

with an intermediate measurement angle (between 0 and $\pi/2$) is absent.

However, as has been pointed out in the Introduction, in [12–14] it was demonstrated with density matrices that the minimum of the conditional entropy is not necessarily located only at the ends of the segment $[0, \pi/2]$; it can also lie inside it. Moreover, a subdomain $Q_\theta$ for the thermal quantum discord for quite a realistic physical model, an XXZ dimer in an exter-nal field, has been recently found (see the preprint [40]). The boundaries of the subdomain $Q_\theta$ are determined from the condition for the splitting of the extremum (minimum) of the conditional entropy at each end of the segment of its domain of definition into two extrema, a maximum and a minimum (bifurcation phenomenon) [40]:

$$S''_{cond}(0) = 0 \tag{49}$$

and

$$S''_{cond}(\pi/2) = 0, \tag{50}$$

where for the second derivatives from (39)–(42) we have

$$S''_{cond}(0) = \frac{1}{4}(a-b+c-d)\left(2\ln\frac{b+d}{a+c} + \ln\frac{ac}{bd}\right)$$
$$+ \frac{1}{4}(a-b-c+d)\ln\frac{ad}{bc} - \frac{1}{2}(|u|+|v|)^2 \tag{51}$$
$$\times \left(\frac{1}{a-c}\ln\frac{a}{c} + \frac{1}{b-d}\ln\frac{b}{d}\right)$$

and

$$S''_{cond}(\pi/2) = (a-b+c-d)^2 - \frac{1}{2(1+r)}\left[a-b+c-d\right.$$
$$\left. + \frac{1}{r}(a+b-c-d)(a-b-c+d)\right]^2$$
$$- \frac{1}{2(1-r)}\left[a-b+c-d\right.$$
$$\left. - \frac{1}{r}(a+b-c-d)(a-b-c+d)\right]^2 \tag{52}$$
$$+ \frac{1}{2r}\left\{(a-b-c+d)^2\left[1 - \frac{1}{r^2}(a+b-c-d)^2\right]\right.$$
$$\left. - 4(|u|+|v|)^2\right\}\ln\frac{1-r}{1+r},$$

here,

$$r = [(a+b-c-d)^2 + 4(|u|+|v|)^2]^{1/2} \tag{53}$$

(this $r$ should not be confused with $r$ from Eqs. (16) for a nanopore either).

Thus, when calculating the quantum discord of an X state, one should use the piecewise-analytic–numerical formula (44) and find the boundaries of the subdomain $Q_\theta$ by solving Eqs. (49)–(53). If the two boundaries coincide, then the subdomain $Q_\theta$ is absent (degenerates to a point) and the piecewise-analytic formula (47) is valid. There can also be the cases where the discord is determined by one branch, $Q_0$ or $Q_{\pi/2}$. Such a situation takes place, for example, for a thermalized Heisenberg dimer in the absence of an external magnetic field [41].

## 7. RESULTS AND DISCUSSION

Applying the general scheme developed above to the problem of a gas in a nanopore, for $Q_0$ and $Q_{\pi/2}$ we find (now in bits to preserve the continuity with the results of our paper [22])

$$Q_0 = -S - \left(\frac{1}{4} + p + q\right)\log_2\left(\frac{1}{4} + p + q\right)$$
$$- 2\left(\frac{1}{4} - q\right)\log_2\left(\frac{1}{4} - q\right) \tag{54}$$
$$- \left(\frac{1}{4} - p + q\right)\log_2\left(\frac{1}{4} - p + q\right),$$

$$Q_{\pi/2} = -S - \left(\frac{1}{2} + p\right)\log_2\left(\frac{1}{2} + p\right) - \left(\frac{1}{2} - p\right)$$
$$\times \log_2\left(\frac{1}{2} - p\right) - D_1\log_2 D_1 - D_2\log_2 D_2, \tag{55}$$

where

$$D_{1,2} = \frac{1}{2}[1 \pm 2[p^2 + (r + |2u\sin 2\varphi - r\cos 2\varphi|)^2]^{1/2}]. \tag{56}$$

In Eqs. (54) and (55), $S$ is the entropy of the density matrix $\rho_{AB}$ for a pair of particles as before but now in a nanopore:

$$S = -\left(\frac{1}{4} + q + \sqrt{p^2 + u^2}\right)\log_2\left(\frac{1}{4} + q + \sqrt{p^2 + u^2}\right)$$
$$- \left(\frac{1}{4} + q - \sqrt{p^2 + u^2}\right)\log_2\left(\frac{1}{4} + q - \sqrt{p^2 + u^2}\right) \tag{57}$$
$$- \left(\frac{1}{4} - q + r\right)\log_2\left(\frac{1}{4} - q + r\right) - \left(\frac{1}{4} - q - r\right)\log_2\left(\frac{1}{4} - q - r\right).$$

Figure 1 shows the behavior of $Q_0$ and $Q_{\pi/2}$ for a nanopore with $N = 10$ and $\beta = 1$ as a function of the dimensionless time $\alpha t$. Both functions are periodic with a period equal to $\pi$. In the interval $0 < \alpha t < \pi$, the curves intersect at the points $\alpha t_1 = 0.98486$ and $\alpha t_2 = 2.15673$. The solution of Eqs. (49)–(53) for the boundaries of the subdomain $Q_\theta$ shows that, in this case, the boundaries coincide between themselves. Consequently, no neighborhoods where the conditional entropy $S_{cond}(\theta)$ would have a minimum inside the interval $(0, \pi/2)$ arise near the intersection of the branches $Q_0(\theta)$ and $Q_{\pi/2}(\theta)$ for the discord in the nanopore. The behavior of the conditional entropy as the instant of time $t_1$ is approached and passed is presented in Fig. 2. It can be seen from the figure that the curve $S_{cond}(\theta)$ begins to flatten out as the point of a sudden transition of the discord from one branch ($Q_{\pi/2}$) to the other ($Q_0$) is approached and becomes a straight line at the instant of time $t_1$. In this case, the optimal measurement angle $\theta$ changes from $\pi/2$ to an arbitrary angle $\theta \in [0, \pi/2]$ (a completely isotropic state). With a further arbitrarily small increase in time, the minimum of the conditional entropy passes to the angle $\theta = 0$ (curve *4* in Fig. 2) and $\sigma_z$ becomes an optimal observable instead of $\sigma_x$. The reverse transition of the optimal measurement angle from $\theta = 0$ to $\theta = \pi/2$ occurs at the instant of time $t_2$ and, subsequently, an alternation of optimal measurements occurs.

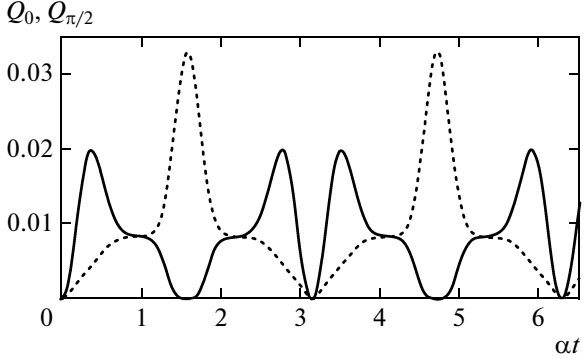

**Fig. 1.** Time dependences for $Q_0$ (solid curve) and $Q_{\pi/2}$ (dashed curve) in a nanopore with $N = 10$ and $\beta = 1$.

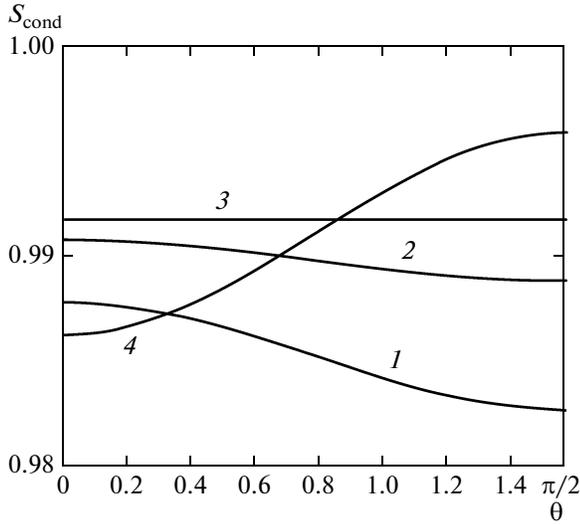

**Fig. 2.** Change in the shape of $S_{\text{cond}}(\theta)$ for a nanopore with $N = 10$ and $\beta = 1$ at the instants of time $\alpha t = 0.6$ (*1*), 0.7 (*2*), 0.98486 (*3*), and 1.3 (*4*).

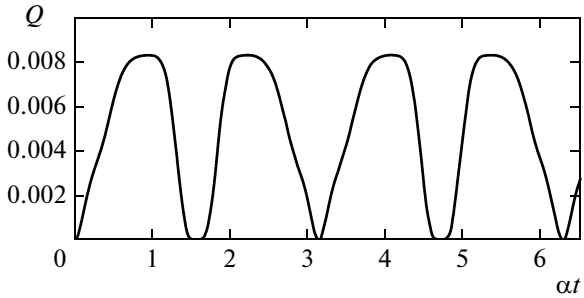

**Fig. 3.** Evolution of the quantum discord in a nanopore with $N = 10$ and $\beta = 1$.

Figure 3 shows the behavior of the quantum discord $Q(\alpha t)$ for the case of a nanopore with $N = 10$ and $\beta = 1$ under discussion. Here, Eq. (47) serves to find the discord. It can be seen from the figure that there are no quantum correlations once a $\pi/2$ pulse has been applied to the system at the initial instant of time $t = 0$ (the state (12) is completely factorized), but they immediately begin to grow, reach a maximum value of 0.008342, and then decay and immediately revive again. The process is repeated. As has been said above, the periodic appearances and disappearances of the quantum discord were called in [22] its flickering. The Fourier expansion of $Q = \alpha t$ gives the flickering spectrum.

Note the following. It follows from Eqs. (16) and the expressions for $Q_0$ and $Q_{\pi/2}$ presented in this section that the quantum discord in the thermodynamic limit $N \longrightarrow \infty$ at instants of time when $Q \neq 0$ is (see also [22])

$$Q = \frac{1}{4}\{(1 + 8q)\log_2(1 + 8q)$$
$$+ (1 - 8q)\log_2(1 - 8q)\} - \frac{1 + 4q}{2}\log_2(1 + 4q) \quad (58)$$
$$- \frac{1 - 4q}{2}\log_2(1 - 4q),$$

with now

$$q = \frac{1}{8}\tanh^2\frac{\beta}{2}. \quad (59)$$

As we see, the discord here depends only on the reduced temperature. Let us set $\beta = 1$. From (58) and (59) we then obtain $Q = 0.0083358...$ Returning again to Fig. 3, we notice that the discord at its maximum is close to this value, slightly exceeding it.

Let us now study the peculiarities of the quantum discord in spin pairs at an odd number of particles in a nanopore. Let $N = 11$ and let the inverse temperature remain $\beta = 1$ as before. Figure 4 illustrates the time variations of $Q_0$ and $Q_{\pi/2}$, candidates for the quantum discord. Both $Q_0$ and $Q_{\pi/2}$ become zero at $\alpha t = \pi l$ ($l = 0, 1, 2, ...$). On the other hand, the curves $Q_0(\alpha t)$ and $Q_{\pi/2}(\alpha t)$ now not only do not intersect but even do not touch each other at the points where $Q_{\pi/2}(\alpha t)$ reaches its largest values. This is shown by an additional analysis with a higher resolution than that provided by Fig. 4. There is no subdomain $Q_\theta$ here again. As a result, the quantum discord at an odd $N$ is $Q = Q_{\pi/2}$, i.e., it is expressed in a closed analytical form just as at an even $N$.

Figure 5 shows the discord dynamics at an odd number of particles in a nanopore. Curiously, although the scheme of discord measurements at even and odd $N$ differ fundamentally from one another (at an even number of particles in the system the measurement angles should be alternated between themselves, while at an odd $N$ the optimal measurement angle always remains the same), the quantum discords at $N$ and $N + 1$ virtually coincide. This can be easily verified by comparing Figs. 3 and 5.

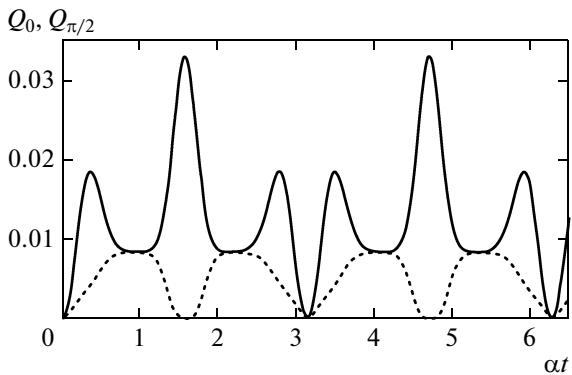

**Fig. 4.** Behavior of $Q_0$ (solid curve) and $Q_{\pi/2}$ (dashed curve) in a nanopore with $N = 11$ and $\beta = 1$.

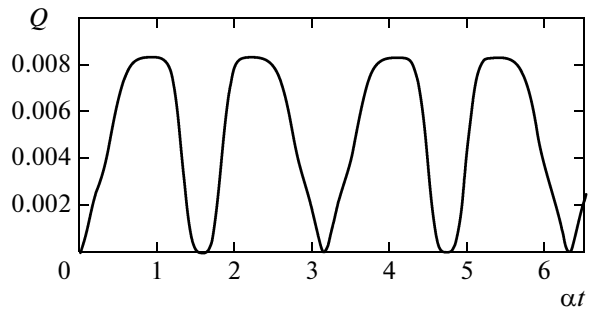

**Fig. 5.** Evolution of the quantum discord in a nanopore with $N = 11$ and $\beta = 1$.

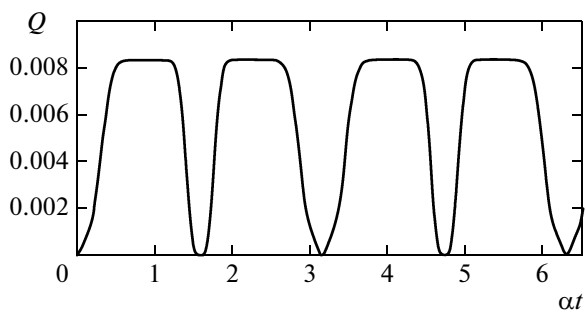

**Fig. 6.** Quantum discord versus time in a nanopore with $N = 20$ and $\beta = 1$.

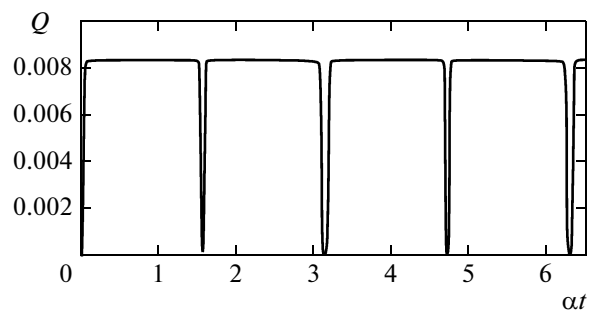

**Fig. 7.** Behavior of the quantum discord in a nanopore with $N = 1000$ and $\beta = 1$.

Let us discuss the behavior of the quantum discord at a large number of molecules in a nanopore. Figures 6 and 7 show the time variations of the discord at $N = 20$ and $10^3$. We see that the maxima of the curves become flat and their magnitudes approach the thermodynamic limit (58) with increasing $N$. In this case, the relative extents of the horizontal plateaus (saturation regions) increase, while the time intervals between them when the correlations drop to zero decrease. In the limit $N \longrightarrow \infty$, we will obviously have a constant quantum discord (58) that, however, periodically completely disappears (fades) for an instant. Note that the period of the function $Q(\alpha t)$ in this limit decreases by a factor of 2: from $\pi$ to $\pi/2$.

## 8. CONCLUSIONS

We studied the dynamics of the quantum-information correlation (discord) for pairs of nuclear 1/2 spins of gas or liquid particles in a closed nanopore in the presence of a strong magnetic field. The dynamics is initiated by applying a resonance $\pi/2$-pulse to the sample in the NMR spectrometer. We showed that correlation oscillations between zero and the maximum value that is determined by the temperature normalized to the external magnetic field strength at a large number $N$ of particles in the nanopore arise in the system thereafter. When $N \longrightarrow \infty$, the oscillations are pulsations (flickering) in character with the complete disappearance of the correlation for a short time and its rapid return to a stationary level. The discord flickering period is determined by the effective spin coupling constant, which, in turn, depends on the size, shape, and orientation of the nanopores in the sample. The discord in the stationary regions is entirely controlled by the normalized temperature. Finally, the ratio of the stationary-region extend to the discord oscillation period depends on $N$. Thus, investigating the behavior of the quantum discord allows certain conclusions about the system's main parameters to be reached.

The physical problem of particles in a nanopore stimulated the development of methods for calculating the quantum discord of CS and X states in the presented paper. These methods can also be applied in other fields, for example, in the quantum information theory.

ACKNOWLEDGMENTS

This work was financially supported by the Russian Foundation for Basic Research (project nos. 13-03-12418, 13-03-00017), Program no. 8 of the Presidium of the Russian Academy of Sciences, and Program no. 14-042 of the Department of Chemistry and Material Science of the Russian Academy of Sciences.